\documentclass[useAMS,usenatbib,usegraphicx]{mn2e}
\usepackage{multirow}
\usepackage{url}

\title[X-ray observations of the PSR~B0540-69]{X-ray observations of the Large Magellanic Cloud pulsar PSR~B0540-69 and its PWN}
   
\author[R. Campana et al.]
  {R.~Campana,$^1$\thanks{E-mail address: \texttt{riccardo.campana@uniroma1.it} }
   T.~Mineo,$^{2}$
   A.~De~Rosa,$^{3}$
   E.~Massaro,$^{1}$
   A.J.~Dean,$^{4}$ 
   L.~Bassani$^{5}$ \\ \\
$^1$ Department of Physics, University of Rome ``La Sapienza'', Piazzale A. Moro 2, I-00185, Rome, Italy \\
$^2$ INAF/IASF-Palermo, Via U. La Malfa 153, I-90146, Palermo, Italy \\
$^3$ INAF/IASF-Roma, Via Fosso del Cavaliere 100, I-00133, Roma, Italy \\
$^4$  School of Physics and Astronomy, University of Southampton, Highfield, SO17 1BJ, United Kingdom \\
$^5$ INAF/IASF-Bologna, Via P. Gobetti 101, I-40129 Bologna, Italy \\
}
      
\date{Accepted 2008 June 18.  Received 2008 June 10; in original form 2008 April 29}
      
\pagerange{\pageref{firstpage}--\pageref{lastpage}} \pubyear{2008}

\begin{document}
\label{firstpage}

\maketitle

\begin{abstract}
PSR~B0540-69 is a young pulsar in the Large Magellanic Cloud that has similar properties with respect to the Crab Pulsar, and is embedded in a Pulsar Wind Nebula. 
We have analyzed the complete archival RXTE dataset of observations of this source, together with new Swift-XRT and INTEGRAL-IBIS data.
Accurate lightcurves are produced in various energy bands between 2 and 60 keV, showing no significant energy variations of the pulse shape. 
The spectral analysis shows that the pulsed spectrum is curved, and is best fitted up to 100 keV by a log-parabolic model: this strengthens the similarities with the Crab pulsar, and is discussed in the light of a phenomenologic multicomponent model.
The total emission from this source is studied, the relative contributions of the pulsar and the PWN emission are derived, and discussed in the context of other INTEGRAL detected pulsar/PWN systems. 
\end{abstract}
	

\section{Introduction}
PSR~B0540-69 is situated in the Large Magellanic Cloud, and is one of the few known extragalactic pulsars. 
Its pulsed emission was discovered by Seward et al. (1984) in the soft X-ray band with the \emph{Einstein} observatory, and thereafter also in the optical (Middleditch \& Pennypacker, 1985) and radio bands (Manchester et al., 1993).

This pulsar is usually referred as the ``Crab Twin'', because of the similar period and period derivative with respect to the well known Crab Pulsar (PSR~B0531+21, see Table \ref{t:crabvs0540} for a comparison). 
However, their pulse shapes are strongly different: 
for the Crab it consists of two peaks, with a phase separation of about 0.4, 
that are visible at all wavelengths, 
while PSR~B0540-69 has a single, broadly symmetric pulse about 0.5 wide in phase, with some structure at the maximum.

\begin{table}
\caption{Comparison between various parameters for Crab Pulsar (PSR~B0531+21) and PSR~B0540-69. 
Data are from the ATNF Pulsar Database, available at 
\texttt{http://www.atnf.csiro.au/research/pulsar/psrcat/}. }\label{t:crabvs0540}
\centering
\begin{tabular}{lcc}
										& Crab 	& PSR~B0540-69 \\ \hline
Period $P$ (ms)							& 33.5	& 50.5 \\
Period derivative $\dot{P}$ ($\times 10^{13}$ s/s)	& 4.2		& 4.8 \\
Braking index $n$ 							& 2.5		& 2.1 \\
Distance $d$ (kpc)							& 2		& 50 \\
Characteristic age $\tau$ (yr)					& 1240	& 1670 \\
Magnetic field $B_{s}$ ($\times 10^{12}$ G)		& 3.78 	& 4.96 \\
Spin-down luminosity ($\times 10^{38}$ erg/s)		& 4.6		& 1.5 \\ 
\hline
\end{tabular}
\end{table}

This pulsar, like the Crab, is also embedded in a Pulsar Wind Nebula (PWN) that shines in the optical to X-ray band (Serafimovich et al., 2004).
This pulsar/PWN  system has been observed many times in the X-rays, by ROSAT (Finley et al., 1993), GINGA (Deeter et al., 1999), BeppoSAX  (Mineo et al., 1999), ASCA (Hirayama et al., 2002), RXTE (de Plaa et al., 2003),  and Chandra (Gotthelf \& Wang, 2000; Hwang et al., 2001; Kaaret et al., 2001; Petre et al., 2007).
This source is not detected in $\gamma$-rays ($>50$ MeV), with only upper limits reported by EGRET (Thompson et al., 1994).

In the optical/near UV band PSR~B0540-69 is a faint $\sim$22 magnitude source with a power law spectrum having a steep spectral index $\alpha_{\nu} \sim 1$ (Serafimovich et al., 2004) while the Crab has a flat spectrum, $\alpha_{\nu} \sim 0$, in the same band.

In the X-rays, this pulsar has a flux of a few milliCrab, and a spectrum fitted by various authors as a power law with photon indices in the range from 1.8 to 2.0. 
De Plaa et al. (2003), using high resolution RXTE spectra, showed evidence for a curved, \emph{log-parabolic} spectrum, i.e. with the form $F(E) = KE^{-a-b\,\mathrm{Log}E}$.
A similar spectrum has been observed also for the Crab pulsar (Massaro et al., 2000, 2006) and few other sources (Cusumano et al., 2001; Mineo et al., 2004).

PSR~B0540-69's  PWN (also called N158A) has a torus-like structure, with a major radius of $\sim$4\arcsec\ and with the presence of jets along the direction of proper motion. 
At variance with the Crab, the PWN of PSR~B0540-69 is further surrounded by a Supernova Remnant (SNR) thermal shell, with a radius of about 30\arcsec, that has radio and soft X-ray emission. Optical (HST; Serafimovich et al., 2004) and X-ray (Chandra; Hwang et al., 2001; Kaaret et al., 2001; Petre et al., 2007) observations showed that the various regions of the PWN/SNR system have different brightnesses and spectra, not described by simple power law models. 

A first analysis of PSR~B0540-69 using INTEGRAL data was performed by Slowikowska et al. (2007), who reported pulse profiles in different energy bands and a total, phase-averaged, spatially-extended spectrum.

In this paper we present a comprehensive X/$\gamma$-ray study of this pulsar-PWN system, analyzing in detail the INTEGRAL observations together with new SWIFT/XRT data. We also exploit the very large RXTE observation dataset of this pulsar, which spans a period of more than ten years.

This paper is structured as follows. In Sect. \ref{obs-datared} we describe the X-ray observations and data reduction and we present our results on the pulsar and PWN spectra. In Sect. \ref{spectra} and \ref{conclusions} we discuss these results and their interpretation also in the context of pulsar-PWN high-energy emission models. 


\section{Data reduction and spectral analysis}\label{obs-datared}

\subsection{Swift XRT}
PSR~B0540-69 was observed fifteen times in three different periods by the XRT (\emph{X-Ray Telescope}) instrument onboard the Swift satellite (Gehrels et al. 2004).
In Table \ref{t:xrtlog} we report the log of these observations, performed between 2005 and 2007.
XRT is a focusing telescope with a CCD detector, and its useful band of sensitivity is roughly 0.7--7 keV. 
The main observing modes are Photon Counting (PC) and Windowed Timing (WT).
In PC mode the data are integrated in a 2.5 seconds exposure, retaining full imaging and spectroscopic capabilities but with very limited time resolution; 
while in WT mode the central 200 CCD columns (8\arcmin\ of the field of view) are continuously registered in one-dimensional mode,  with a time resolution of 1.7 ms.

\begin{table}
\caption{Log of XRT observations of PSR~B0540-69.}\label{t:xrtlog}
\centering
\begin{tabular}{c cccc}
No. & ObsID &	Start date & Mode & Exposure (s) \\ \hline
01 & 00053400001 & 2005 Apr 14 & PC & 8\,415 \\
02 & 00053400002 & 2005 Apr 19 & PC & 3\,319 \\
03 & 00053400003 & 2005 Apr 20 & PC & 6\,644 \\
04 & 00053400004 & 2005 Apr 28 & PC & 7\,340 \\
05 & 00053400005 & 2005 May 04 & PC & 8\,071 \\ \hline
06 & 00053402001 & 2006 Jan 26 & WT & 2\,815 \\
07 & 00053402002 & 2006 Feb 28 & WT & 1\,505 \\
08 & 00053402003 & 2006 Mar 04 & WT & 16\,548 \\
09a & 00053402004 & 2006 Apr 18 & WT & 2\,651 \\
09b &		''	 &	''	         & PC & 3\,866 \\
10 & 00053402005 & 2006 Apr 19 & PC & 11\,205 \\
11 & 00053402006 & 2006 Apr 20 & PC & 6\,261 \\ \hline
12 & 00053402007 & 2007 Nov 05 & PC & 5\,147 \\
13 & 00053402008 & 2007 Nov 08 & PC & 3\,260 \\
14 & 00053402009 & 2007 Nov 09 & PC & 11\,431 \\
15 & 00053402010 & 2007 Nov 11 & PC & 15\,438 \\
\hline
\end{tabular}
\end{table}

Raw event data (level 1) were reduced following the standard pipeline (\texttt{xrtpipeline} v.0.11.6 task) using the HEADAS v.6.4 package, including SWIFT software v.2.8 and the latest CALDB calibration files. Events were screened adopting standard grade filtering.
The total observation time is respectively 90.4 ks for the PC mode and 23.5 ks for WT mode. We used the PC mode observations to obtain a total (pulsar+PWN) source spectrum, and the WT mode observation to derive the pulsed spectrum. 

\subsubsection{Photon Counting mode observations}
Source spectra were extracted from a circular region of radius 20 pixels (equal to 47\arcsec .2) excluding the central pixel to avoid count rates higher than 0.6 cts/s and thus pile-up contamination.
The PWN radius ($\sim$4\arcsec) is comparable to the XRT point spread function, so this source is almost point-like.
Background spectra were extracted from an annular region outside the source region, with radii of 20 and 50 pixels respectively.
The corresponding Ancillary Response Files (ARF) were generated by means of the \texttt{xrtmkarf} task, taking into account the exposure map generated by the pipeline.

\begin{figure}
\centering
\includegraphics[angle=270,scale=0.35]{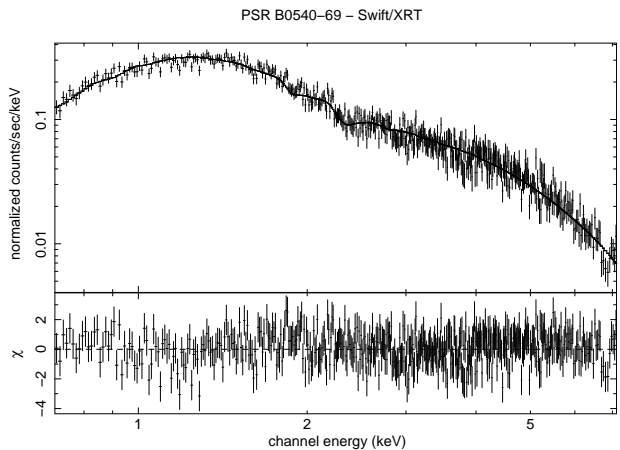}  
\caption{Swift XRT X-ray (0.7--7.14 keV) spectrum of  the total (pulsar+PWN) emission from PSR~B0540-69.}\label{f:total}
\end{figure}

All spectra, rebinned to have at least 20 counts for each energy channel, were fitted individually with XSPEC v. 11.3.1, using an absorbed powerlaw (\texttt{wabs*powerlaw}) model. 
The results  were verified to be consistent, and all were found to match within the errors: the spectra were then summed together.
The result is shown in Fig. \ref{f:total}, and the best fit model has an hydrogen column density $N_{H} =  (3.7 \pm 0.1) \cdot 10^{21}$ cm$^{-2}$ and a photon index $\Gamma = 1.98 \pm 0.02$. The reduced $\chi^{2}$ is 1.01 for 483 degrees of freedom.
The 2--10 keV absorbed flux is $(2.61\pm0.03)\cdot10^{-11}$ erg~cm$^{-2}$~s$^{-1}$.

It should be noted that our fit assumes Galactic elemental abundancies and doesn't discriminate between Milky Way and LMC absorption, although the former is about one order of magnitude less than the latter. 
The derived $N_{H}$ value, therefore, should be regarded as a simple, phenomenological parameter for the description of the spectrum. 
See Serafimovich et al. (2004) for an extensive discussion on the photoelectric absorption in the X-ray spectrum of PSR~B0540-69.

Our results are in very good agreement with previous observations: BeppoSAX LECS and MECS (Mineo et al. 1999) data gave in fact $N_{H} =  (3.7 \pm 0.1) \cdot 10^{21}$ cm$^{-2}$ and $\Gamma = 1.94 \pm 0.03$, also in agreement with previous ROSAT results (Finley et al., 1993).
Moreover, summing the Chandra pulsar phase-averaged and PWN spectra determined by Kaaret et al. (2001), we also obtain a very good agreement with our XRT total spectrum (see Figure \ref{f:sed}).

We also evaluated what is the contribution of the SNR thermal shell observed by Chandra on the XRT spectrum. From Hwang et al. (2001) it is apparent that, at 20\arcsec--40\arcsec\ from the pulsar, the SNR falls within XRT spectral extraction radius. However, the thermal shell is very soft, with almost all of the emission below 2 keV. 
Moreover, from the analysis of Petre et al. (2007), the flux from the shell is only a few percent of the total PWN emission.
Therefore, the SNR contamination on the XRT spectral fit is negligible.

\subsubsection{Windowed Timing mode observations}\label{xrt-wt}
In order to obtain the pulse profile and phase resolved spectra, the photon arrival times of the four WT-mode observations (obs. \# 06, 07, 08 and 09a, see Table \ref{t:xrtlog}), 
selecting the events in a 40 pixel-wide box centered on the pulsar position, were reduced to the solar system barycentre, using JE200 ephemeris and the spacecraft orbital data. 

During the observation \# 08 the source was erratically inside and outside the field of view, therefore considerably reducing the useful exposure: we selected only the periods with a count rate higher than 0.5 cts/s, obtaining a total net exposure time shorter than 2 ks.
 
\begin{figure}
\centering
\includegraphics[angle=270,scale=0.35]{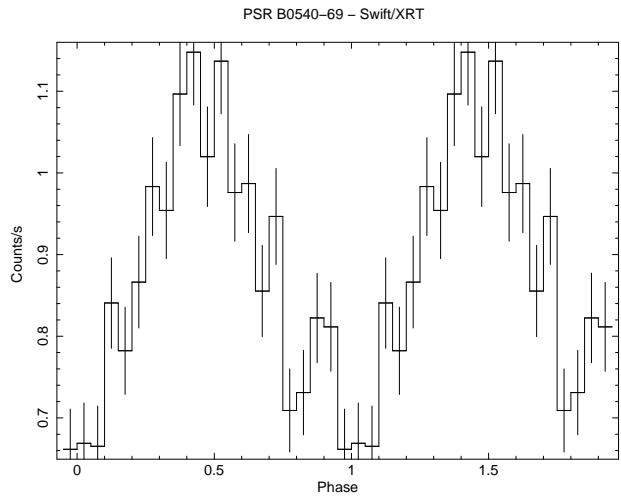} 
\caption{Folded lightcurve in 20 phase bins for XRT observations of PSR~B0540-69, in the energy range from 0.7 to 7.14 keV.}\label{f:xrt_lc}
\end{figure}

Radio timing observations spanning the XRT data epoch are not available, 
therefore we performed a period search, folding the light curves at various trial frequencies 
and computing the corresponding $\chi^{2}$ value with respect to a constant level. 
Significant pulsation was found only for observations \# 05 and 09a, the longest ones.
The frequency corresponding to the maximum $\chi^{2}$ value was thus obtained (Table \ref{t:xrteph}), and the folded lightcurves were arbitrarily shifted to have the pulse maximum (determined by a Gaussian fit) at the phase 0.5.
The curves were then added together. 
The resulting total light curve, for a net exposure time of 5472 seconds, is shown in Figure \ref{f:xrt_lc}.

\begin{table}
\caption{Swift-XRT best fit folding frequencies.}\label{t:xrteph}
\centering
\begin{tabular}{ccc}
Obs. \# & $\nu_{0}$ (Hz) &	Reference epoch (MJD)  \\ \hline
06 & 19.7609226 & 53761.762  \\
09a & 19.7595952 & 53843.561 \\ \hline
\end{tabular}
\end{table}

A phase-averaged spectrum was computed and yielded results that were perfectly consistent with the PC-mode observations. 
We assumed the phases 0.90--1.10 as the off-pulse, representative of the nebular spectrum, 
while 0.10--0.90 as the pulsar phase.
We subtracted the off-pulse from the pulsed data, obtaining a background-subtracted pulsed spectrum, that was fitted with an absorbed power-law, using the $N_{H}$ value derived from the total emission analysis, obtaining $\Gamma = 1.6 \pm 0.2$ and a 2--10 keV absorbed flux of $(6.4\pm0.8)\cdot10^{-12}$  erg~cm$^{-2}$~s$^{-1}$, in agreement with the previous BeppoSAX observations (Mineo et al., 1999).

\subsection{INTEGRAL IBIS/ISGRI} \label{ibis}
The INTEGRAL (International Gamma Ray Astrophysics Laboratory) mission observed the Large Magellanic Cloud during nine orbits in two periods,  January 2003 and January 2004 (Gotz et al. 2006).
This satellite has two main instruments, the imager IBIS and the spectrometer SPI.
IBIS (\emph{Imager on Board Integral Satellite}, Ubertini et al. 2003) is a coded-mask imager and is constitued by two simultaneously operating detector layers, ISGRI and PICsIT. 
ISGRI (\emph{Integral Soft Gamma-Ray Imager}, Lebrun et al. 2003) is an array of 8 CdTe solid state detection units, each having 32$\times$64 pixels, thus making an overall 128$\times$128 array, that has a 12\arcmin\ (FWHM) image resolution and a nominal operating range from 15 keV to 1 MeV.
Each INTEGRAL orbit lasts about three days and consists of several pointings, because of the ``dithering'' observational mode. Each pointing, the duration of which is typically half an hour, forms a ``Science Window'' (ScW).

PSR~B0540-69 was observed during revolutions 27, 28, 29, 33, 34, 35 (January 2003) and 150, 151, 152 (January 2004). 
Individual pointings for all the observations performed by the first 3.5 years of INTEGRAL observations were processed using the standard software package OSA v.5.1 (Goldwurm et al., 2003) and mosaicked to obtain an all-sky image. PSR~B0540-69 is then detected at $6.5\sigma$ significance for about 221 ks of corrected on-source exposure (see Bird et al., 2007, for the details of data analysis).
The spectrum is obtained calculating the mean flux in various energy bands spanning the range 20--100 keV.

\begin{figure}
\centering
\includegraphics[angle=270,scale=0.35]{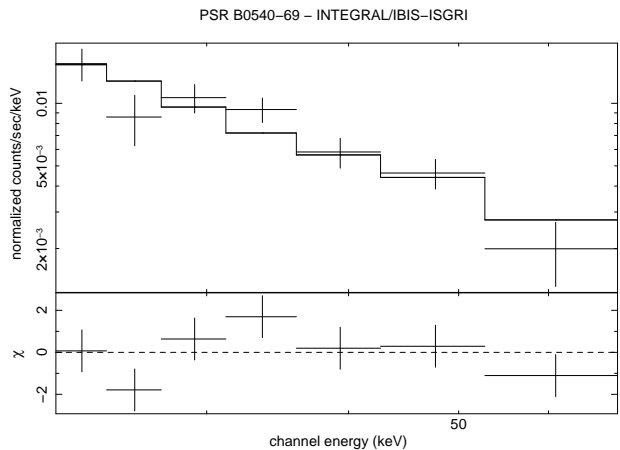} 
\caption{INTEGRAL IBIS-ISGRI (20--100 keV) spectrum of  PSR~B0540-69.}\label{f:ibis}
\end{figure}

A power-law fit of IBIS data in the 20--100 keV band is reported in Fig. \ref{f:ibis}. The photon index is $\Gamma = 2.12 \pm 0.25$ and the reduced $\chi^{2}$ is 1.5 (for 5 d.o.f.).
The 20--100 keV flux is $2.9^{+0.1}_{-1.2}\cdot10^{-11}$  erg~cm$^{-2}$~s$^{-1}$, in agreement with Gotz et al. (2006).
 
\begin{figure}
\centering
\includegraphics[angle=270,scale=0.35]{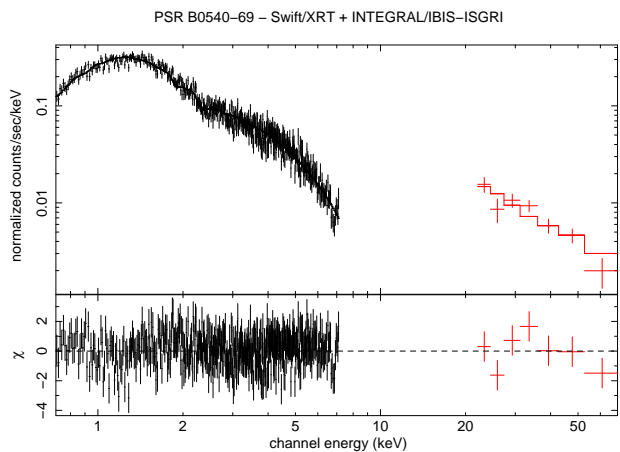} 
\caption{Swift XRT  (0.7--7.14 keV) and INTEGRAL IBIS-ISGRI (20--100 keV) spectrum of  PSR~B0540-69.}\label{f:simultaneo}
\end{figure}

Note that this photon index is steeper than the XRT one, although compatible because of the large uncertainties.
In fact, a simultaneous fit of XRT and IBIS data for the total source (PSR+PWN) spectrum is reported in Fig. \ref{f:simultaneo}. The resulting parameters are: $N_{H} =  (3.8 \pm 0.1) \cdot 10^{21}$ cm$^{-2}$, $\Gamma = 1.97 \pm 0.02$ and $\chi^{2} = 1.09$/489 d.o.f.
The intercalibration factor between IBIS and XRT data is $f = 1.05 \pm 0.08$, and clearly the fit is constrained by the high-quality XRT data.
The IBIS spectrum thus can be regarded as a smooth extrapolation of the XRT data.

\subsection{RXTE}
\subsubsection{PCA}\label{pca}
PSR~B0540-69 was observed several times with the Rossi X-Ray Timing Explorer (RXTE) satellite. 
We used data obtained with the Proportional Counter Array (PCA, Jahoda et al., 1996) instrument in the Good Xenon mode, that ensures the full timing accuracy (about 1 $\mu$s).
PCA is a $1^{\circ}\times 1^{\circ}$ collimated array of proportional counters, and is composed of 5 PCUs (Proportional Counter Units). The useful sensitivity range is about 2--60 keV.

\begin{table*}
\caption{Log of RXTE observations of PSR~B0540-69. Each proposal ID contains several observations (from 2 to 95). The offset is between the center of the pointing field of view and the location of the pulsar. The on-source exposure time is after the screening (PCA) and dead-time (HEXTE) corrections.}\label{t:rxtelog}
\centering
\begin{tabular}{ccccccc}
Proposal 	& First--Last & Offset			& PCA 			& HEXTE-A 	&  HEXTE-B \\ 
ID 	& Observation Date &  (arcmin) 	&  exposure (s) 	& exposure (s) 	&  exposure (s) \\ \hline
10206 &  1996 Aug 11 -- 1996 Nov 17 	& 0.30  & 54\,049 	& 35\,200	 	& 34\,955	 	\\
10218 &  1996 Oct 12 -- 1996 Dec 22  	& 25.0  & 92\,086 	& 29\,306	 	& 29\,212	 	\\
10250 &  1996 Feb 02 -- 1996 Oct 04  	& 24.6  & 37\,837 	& 12\,422 		& 12\,325	 	\\
20188 &  1996 Nov 30 -- 1997 Dec 12 	& 24.5  & 301\,275 	& 118\,935 	& 118\,114 	\\
30087 &  1998 Jan 04 -- 1998 Sep 30 	& 24.6  & 103\,455 	& 44\,566	 	& 44\,265	 	\\
40139 &  1999 Jan 19 -- 2000 Feb 15 	& 15.8  & 185\,520 	& 71\,834	 	& 71\,439	 	\\
50103 &  2000 Mar 13 -- 2001 Mar 15 	& 15.8  & 219\,360 	& 178\,802 	& 177\,325 	\\
50414 &  2000 Jun 05 -- 2000 Jun 23 	& 0.30  & 5\,552	& 3\,634	 	& 3\,601	 	\\
60082 &  2001 Mar 26 -- 2003 Feb 21 	& 15.8  & 281\,719 	& 223\,510 	& 221\,934 	\\
70092 &  2002 Mar 21 -- 2003 Mar 30 	& 15.8  & 268\,534 	& 233\,883 	& 232\,737 	\\
80089 &  2003 Apr 16 -- 2004 Oct 10 	& 15.8  & 263\,151	& 211\,649 	& 210\,143 	\\
80118 &  2004 Jan 07 --  2004 Jan 12 	& 24.6  & 40\,977 	& 21\,593		& 21\,606	 	\\
90075 &  2004 Mar 03 -- 2005 Mar 29 	& 15.8  & 83\,507 	& 67\,633	 	& 75\,325		\\
91060 &  2005 Jun 20 -- 2006 May 28 	& 15.8  & 47\,523 	& 32\,707	 	& 39\,855	 	\\
92010 &  2006 Apr 25 -- 2007 Jun 22 	& 15.8  & 33\,145	& 27\,210	 	& 29\,755	 	\\
\hline
Total	 (Ms)	 & --- 	    				& --- 	     & 2.018 	& 1.313 		& 1.322 \\
\hline
\end{tabular}
\end{table*}

\begin{figure}
\centering
\includegraphics[angle=270,scale=0.3]{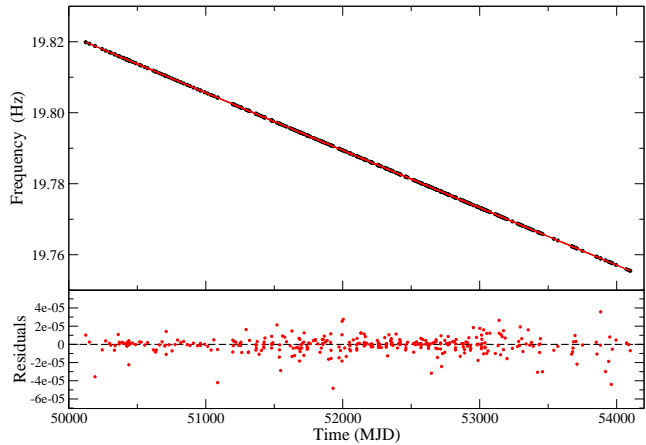} 
\caption{Eleven years of PSR~B0540-69 RXTE-PCA timing. In the lower panel are reported the residuals after the fit with Eq. [\ref{e:timing}], which is shown as the red continuous line in the upper panel.}\label{f:timing}
\end{figure}

\begin{figure}
\centering
\includegraphics[angle=270,scale=0.35]{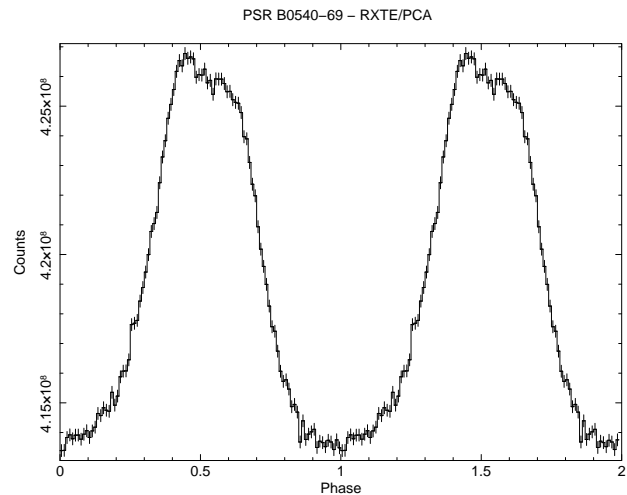} 
\caption{Total light curve obtained from the full PCA dataset, in the energy range 2--60 keV, with 100 phase bins.}\label{f:lc-100}
\end{figure}

\begin{figure*}
\centering
\includegraphics[angle=270,width=\textwidth]{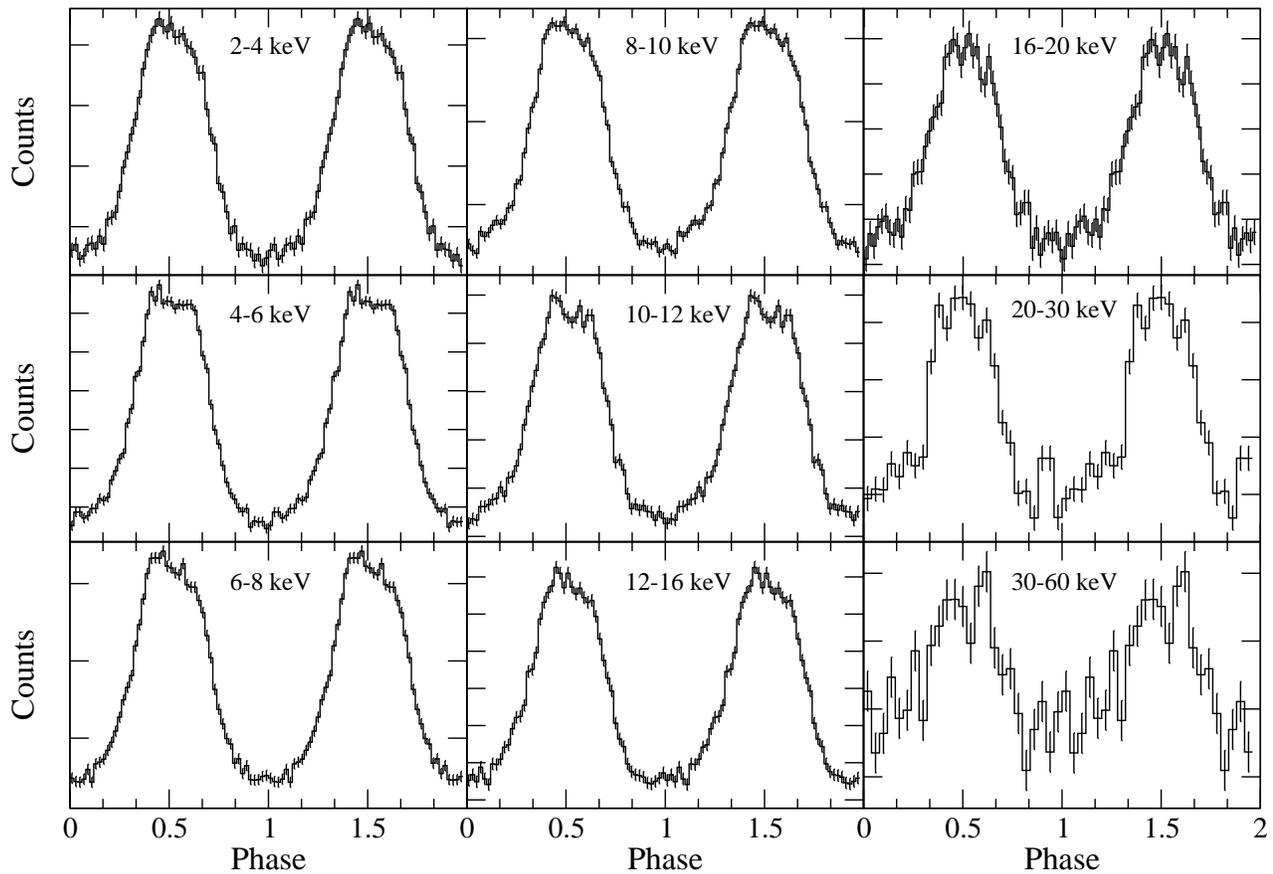} 
\caption{Light curves in various PCA energy bands, from 2 to 60 keV. The curves are with 50 phase bins, except for the last two that have 25 bins.}\label{f:lc-multiplo}
\end{figure*}

For the timing analysis we selected all the archive observations having PSR~B0540-69 at less than 30\arcmin\ from the centre of the field of view (see Table \ref{t:rxtelog}) and with a duration longer than 3000 seconds.
The data set spans a 11 years time range, from 1996 to 2007.
The observations were screened for South Atlantic Anomaly passages, elevation angle and electron background, following the standard criteria, and barycentred using the \texttt{faxbary} task of the FTOOLS package.
The total screened exposure time is more than 2 Ms,
a significant improvement, by a factor of $\sim$3, with respect to the previous RXTE analysis of de~Plaa et al. (2003), which had a total exposure of 684 ks.
A period search was performed for each observation, and the found frequencies (Figure \ref{f:timing}) were fitted with a second-order Taylor expansion of the frequency history,
\begin{equation}\label{e:timing}
\nu(t) = \nu_{0} + \dot{\nu}_{0}(t-t_{0}) + \frac{\ddot{\nu}_{0}}{2} (t-t_{0})^{2}
\end{equation}
 in order to find a satisfactory ephemeris spanning the data set.
The values derived from the fit are, for the time range in MJD 50100--54100: $\nu_{0} =  19.774$ Hz, $\dot{\nu}_{0} =  -1.87253 \cdot10^{-10}$ Hz/s and $\ddot{\nu}_{0} = 3.69 \cdot 10^{-21}$ Hz/s$^{2}$, for the reference epoch $t_{0}$  MJD 52954.
In the 11-years time interval of RXTE observations the spin frequency of the pulsar decreases from a value of about 19.82 Hz to 19.75 Hz.
Note that the XRT folding frequencies (Sect. \ref{xrt-wt}) are in good agreement with the extrapolation of these ephemeris at the relative epochs.
The values of the frequency and its time derivative are in good agreement with other works (Cusumano et al., 2003; Livingstone et al., 2005; Slowikowska et al., 2007). However, the value of the second time derivative is not well constrained by the fit. 
Each observation was thus phase-folded. The alignment between the pulse profiles of each observation was performed by a cross-correlation method, first with a sinusoidal function in order to have the peak maximum at phase 0.5, then with an analytic template based on the resulting total light curve.
A 100-bin pulse profile in the energy band 2--60 keV is showed in Figure \ref{f:lc-100}, and energy-resolved profiles are showed in Figure \ref{f:lc-multiplo}.

From this figure it is apparent that there is no evident change of the light curve shape with energy. Following de Plaa et al. (2003), we fitted the full pulse profile (Figure \ref{f:lc-100}) with an analytical model described by the sum of two Gaussians and a constant. We then performed this fit on all the Figure \ref{f:lc-multiplo} light curves, fixing the central phase and the widths of the Gaussians at the values determined in the total light curve fit, and we found no statistical significant variation on the relative normalizations of the Gaussians. 
We also checked for the stability of the pulse profile, comparing it as observed at various epochs, and we did not found any significant changes in shape or flux.

\begin{figure*}
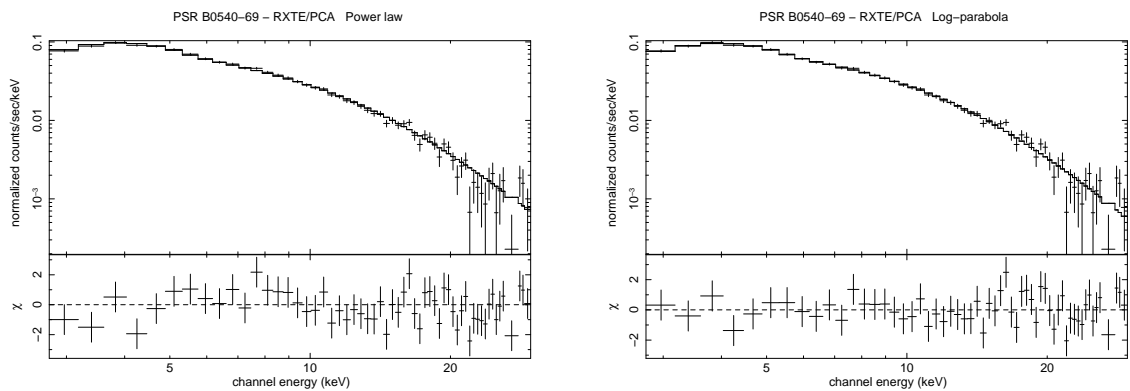

\centering
\includegraphics[angle=270,scale=0.3]{fig8a}\hspace{1cm}
\includegraphics[angle=270,scale=0.3]{fig8b}
\caption{\emph{Left}. Power-law fit of the 2--30 keV PCA pulsed spectrum. Note the systematic, curved behaviour of the residuals. \emph{Right}. Log-parabolic fit of the spectrum in the same energy range.}\label{f:pcaspectra}
\end{figure*}

For the spectral extraction we used only the observations pertaining to the last gain epoch (Epoch 5, from May 13 2000 onward, proposal ID 50103 to 92010) in order to ensure the consistency of the analysis.
The resulting dataset has a total exposure of 1.2 Ms.
In a similar way with respect to XRT, the phases 0.10--0.90 were selected as the pulsed signal, and the phases 0.90--1.10 as the offpulse, that was subtracted on the pulsed signal.
Offset-corrected response matrices were determined for each observation and each PCU, and added together weighting them by their exposure. 
Spectra were also extracted for each PCU, rebinned to have a minimum of 20 counts for channel (however, the rebinning affects only the high energy channels, over 30 keV), and verified for consistency.

A power-law fit of the pulsed signal, in the energy range 2--30 keV 
and with $N_{H}$ frozen at the value given by the XRT analysis, 
gave the values $\Gamma = 1.89 \pm 0.01$ and $\chi^{2} = 1.14$/59 d.o.f.; 
the residuals, however, showed a systematic behaviour typical of curved spectra, with an excess in the central bins
(see Figure \ref{f:pcaspectra}, left panel).
We considered therefore a log-parabolic law, and the fit gave the values $a = 1.47 \pm 0.01$, $b = 0.25 \pm 0.06$ and $\chi^{2} = 0.96$/58 d.o.f., and the residuals don't show any systematic deviation (Figure \ref{f:pcaspectra}, right panel).
We performed an $F$-test in order to assess the statistical significance of the log-parabolic model versus the power law, and the null hypothesis probability resulted $9.7\cdot10^{-4}$.
The 2--10 keV pulsed absorbed flux is $6.5^{+0.4}_{-0.7}\cdot10^{-12}$ erg~cm$^{-2}$~s$^{-1}$, in agreement with the XRT value.

The 30--60 keV PCA spectrum has a much lower statistics. Extending the previous fit up to 60 keV does not alter the values of the parameters. The 30--60 keV spectrum alone is well described by a single power law with a photon index around 2.

There are some discrepancies with respect to the spectral results of de~Plaa et~al. (2003), both in the curvature parameter and in flux. 
Firstly, they gave the log-parabolic law in terms of \emph{natural}, instead of decimal, logarithm. However, also correcting for the change of logarithm basis, their curvature parameter is larger (i.e. the log-parabola is more curved) than ours. 
There are many possible explanations for the discrepancy: 
(\textit{a}) the different methods used to determinate the pulsed spectrum (de~Plaa et~al. determined the pulsed counts excess in 14 energy bins by means of an analytical fit of the light curve); 
(\textit{b}) the different energy range of their fit (that included also soft X-ray ROSAT data), together with a different $N_{H}$ value, could affect the curvature;  
(\textit{c}) their spectral PCA dataset has an exposure (462 ks) much smaller than ours (1.2 Ms); 
and (\textit{d}) a real change in the spectrum between the calibration epoch 3 (1996-99) dataset used by de Plaa et al. and our epoch 5 dataset (2000-07).
Regarding the first of the above points, we determined the excess counts in various energy bands by means of analytical fits and we found them in agreement with the pulsed counts determined by our offpulse-subtraction method.
We also performed a spectral analysis of the epoch 3 data (proposal ID 10206 to 30087) 
with our approach, and we found that both the spectral parameters and the flux are perfectly 1$\sigma$ compatible with the better statistics of the epoch 5 spectrum.
Moreover, the flux values of de Plaa et al. are greater than ours by a factor of about 30\%. 
This discrepancy,
too great to be due only to the different estimation of the pulsed counts above the offpulse level,
could be due to the different way of estimating the effective exposure time.
We verified, in fact, that our data agree very well with the ones of de Plaa et al. when the exposure is reduced by a factor equal to the pulsed phase fraction.
Note that the Chandra pulsed spectrum  in the range 0.6--10 keV (Kaaret et al., 2001, also reported by Serafimovich et al., 2004) matches our data, as seen in Figure \ref{f:sed}. 

\subsubsection{HEXTE}

Data from the instrument HEXTE (High Energy Timing Experiment, Rothschild et al., 1998) onboard RXTE were also used. HEXTE employs NaI(Tl)--CsI(Na) phoswich scintillation detectors, and consists of two separated clusters (A/0 and B/1), each containing four detectors. Detector 2 of cluster B has lost its spectral capabilities, and is useful only for timing analyses. 
We extracted the data using the science event mode that ensures the full timing resolution (8 $\mu$s), and we barycentered and folded the data using the PCA ephemeris. 
The total on-source, dead-time corrected exposure is about 1.3 Ms for each cluster.

\begin{figure}
\centering
\includegraphics[angle=270,scale=0.35]{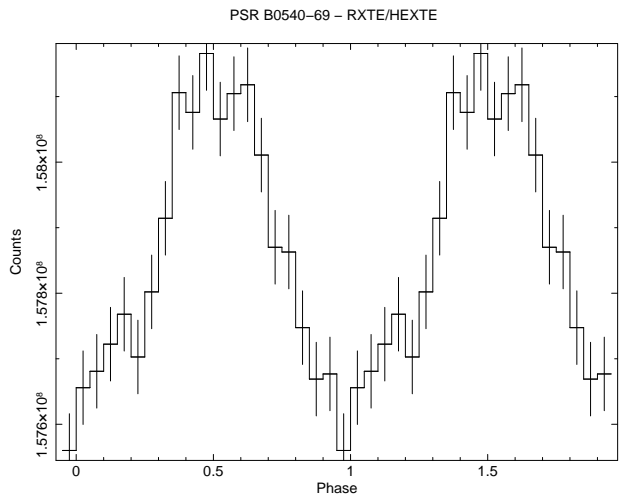} 
\caption{Total light curve obtained from the HEXTE dataset, in the energy range 15--100 keV, with 20 phase bins.}\label{f:lc_hexte}
\end{figure}

Phase alignment was performed using the shift determined from the simultaneous PCA observations.
The folded lightcurve is showed in Figure \ref{f:lc_hexte}.
To increase the signal to noise ratio for the spectral extraction, we used only the observations having an offset less than 16\arcmin, being the response dramatically dependent on the offset angle. 
We produced the offset-corrected response matrices, using the task \texttt{hxtrsp}, and added them together weighting them by the exposure, in a similar way to what was done for PCA. The pulsed spectrum was extracted in the same phase bins of PCA.
In the 15--250 keV range it is well approximated by a single power law with a photon index of $2.1\pm0.2$.
The 20--100 keV pulsed flux is $6.1^{+0.6}_{-2.1}\cdot10^{-12}$ erg~cm$^{-2}$~s$^{-1}$.
Note that this photon index, although affected by a rather large uncertainty, is higher than the one measured in the keV range, and shows a high energy spectral steepening.
We see (Fig. \ref{f:sed}), in fact, that the HEXTE spectrum is fully compatible with the log-parabolic fit of PCA data, and follows smoothly the curvature up to 100 keV. HEXTE data, on the contrary, would lie systematically below the extrapolation of the power law fit of PCA data.

\section{The X-ray spectra of the pulsar and the PWN}\label{spectra}

\begin{table*}
\caption{Results of the spectral analysis. The absorbed flux values are in units of $10^{-11}$ erg~cm$^{-2}$~s$^{-1}$.}\label{t:fitresults}
\centering
\begin{tabular}{c|cccc}
Instrument &  Parameter  & Flux (2--10 keV)  & Flux (20--100 keV)  \\
\hline
Swift/XRT - Total & $\Gamma = 1.98 \pm 0.02 $ & $2.61 \pm 0.03$ & --- \\ 
INTEGRAL/IBIS - Total& $\Gamma = 2.12 \pm 0.25 $ & --- & $2.9^{+0.1}_{-1.2}$ \\ 
Swift/XRT - Pulsed & $\Gamma = 1.6 \pm 0.2$ & $ 0.64 \pm 0.08$ & --- \\ 
\multirow{2}{*}{RXTE/PCA - Pulsed} & $a = 1.47 \pm 0.01$ & \multirow{2}{*}{$0.65^{+0.04}_{-0.07}$} & \multirow{2}{*}{---} \\
 & $b = 0.25 \pm 0.06$ &  &  \\
RXTE/HEXTE - Pulsed & $\Gamma = 2.1\pm0.2$ & --- & $0.61^{+0.06}_{-0.21}$ \\ 
\hline
\end{tabular}
\end{table*}

\begin{figure*}
\centering
\includegraphics[angle=270,width=0.8\textwidth]{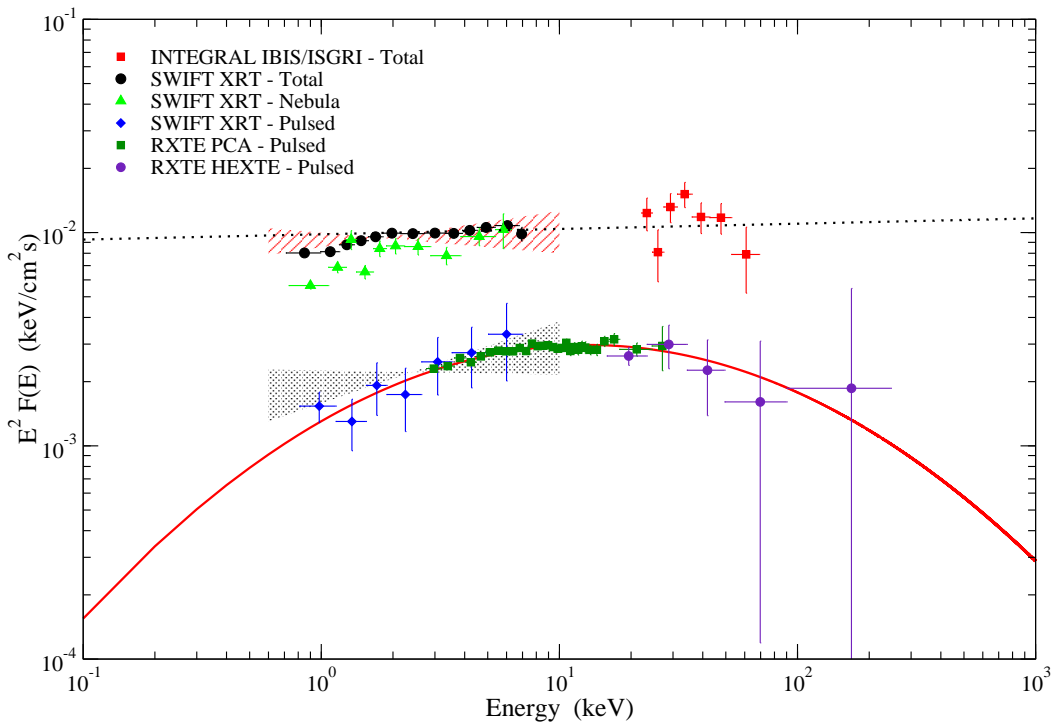} 
\caption{Spectral Energy Distribution (SED) for the emission from PSR~B0540-69 and its PWN. The red continuous line is the log-parabolic law that fits PCA pulsed data (see Sect. \ref{pca}), while the black dotted line  is the power-law simultaneous fit of XRT and IBIS total data (see Sect. \ref{ibis}). The black dotted and red dashed areas are the $\pm 1 \sigma$ Chandra pulsed and total (pulsar phase averaged + PWN) data, respectively (from Kaaret et al., 2001, and Serafimovich et al., 2004). }\label{f:sed}
\end{figure*}

The spectral analysis of the total (pulsar + PWN) and pulsed emission from PSR~B0540-69, described in the previous Section, is summarized in Table \ref{t:fitresults}.
The Spectral Energy Distributions of the total and pulsed emission are plotted in Figure \ref{f:sed}.

The agreement between the various instruments in the 1--10 keV band is very good, both for the total emission and the pulsed one.

It's clearly visible how the pulsed emission accounts only for about 20\% of the total emission in the soft and hard X-rays.
The ratio of the total source flux to the pulsed flux is in fact $\sim5$ in the 20--100 keV band (ISGRI total versus HEXTE pulsed spectrum) and $\sim4$ in the 2--10 keV band (XRT total versus PCA pulsed spectrum).

Therefore, subtracting the pulsed flux from the total one to evaluate the PWN emission in the range 1--100 keV, we obtain an energy-dependent flux ratio $F_{\mathrm{PWN}}/F_{\mathrm{PSR}}$ between 2.5 around the peak of the pulsed emission at about 12 keV, and $\sim$6 at the range boundary. 
Integrating over the whole energy range, we derive a luminosity ratio $L_{\mathrm{PWN}}/L_{\mathrm{PSR}} \sim 3.6 $.
This value is expected to correspond also to the ratio between the efficiencies $\eta_{\mathrm{PWN}}/\eta_{\mathrm{PSR}}$ of conversion of the spin-down luminosity $\dot{E}$ to electromagnetic radiation, and is in agreement with the mean value ($\sim$4) for a large sample of X-ray pulsars (Kargaltsev \& Pavlov, 2008)

For the total energy conversion efficiency, using $\dot{E} =  1.5\cdot10^{38}$ erg/s, we obtain a value of $\eta_{\mathrm{TOT}} = 5.8$\% for the 20--100 keV energy range. The PWN emission contributes to this figure for the 4.6\%, while the pulsed emission for the 1.2\%. Moreover, we have $\eta_{\mathrm{TOT}} = 5.2$\% for the 2--10 keV energy range (PWN: 3.9\%, pulsar 1.3\%).

If we subtract the log-parabolic, pulsed spectrum to the power-law, total spectrum, we can derive an approximate photon index for the nebular spectrum. For the 2--10 keV band we obtain $\Gamma \simeq 2.05$, slightly softer than that of the total emission, in agreement with the mean photon index of Chandra nebular and offpulse spectra ($\sim 2.04$, Kaaret et al., 2001).

Petre et al. (2007) have shown that in the 1--10 keV band the PWN photon index increases from a value of $\sim$1.4 near the pulsar, to $\sim$2.5 at a radius of 4\arcsec. 
The nebular size, however, shrinks with the energy. 
An extrapolation of the relation found by Petre et al. to the IBIS energy range results in a PWN radius of about 2.5\arcsec: inside this radius, the mean photon index is about 1.8.
Again, this is consistent with the derived PWN spectrum obtained subtracting the pulsed contribution to the total spectrum.
For the 20--100 keV band, being after the peak of the pulsed spectrum, the PWN has an harder spectrum than the total emission, with a photon index around 1.8.
This qualitative agreement, however, relies on the assumptions that both the nebular photon indices and the parameterization of the nebular radius could be extrapolated to the hard X-rays, and of course it is affected by the low statistics data for both the total and the pulsed emission.

It is interesting to compare the soft $\gamma$-ray properties of PSR~B0540-69 with the other INTEGRAL-detected pulsar/PWN systems
(PSR~J1846-0258, McBride et al., 2008; PSR~J1617-5055, Landi et al., 2007; Vela, Hoffmann et al., 2006; PSR~J1513-5906, Forot et al., 2006; PSR~J1811-1925, Dean et al., 2008). 
With $\eta\simeq5.8$\% for the total emission, this source has the highest conversion efficiency: the mean of the sample is around 1\%, while Vela is a notable exception with $\eta\simeq0.02$\%. 
Moreover, the $\sim$2.1 photon index of PSR~B0540-69 fits squarely in the observed range for this sample of IBIS sources (2.0--2.3).
The majority of these systems have the hard X-ray emission dominated by the PWN, except for PSR~J1617-5055 and PSR~J1513-5906. The relative contribution from the pulsar and the PWN, however, differs very much from source to source.
A more detailed comparison will be the subject of a future paper (Dean et al., in preparation).

\begin{figure*}
\centering
\includegraphics[angle=270,width=0.8\textwidth]{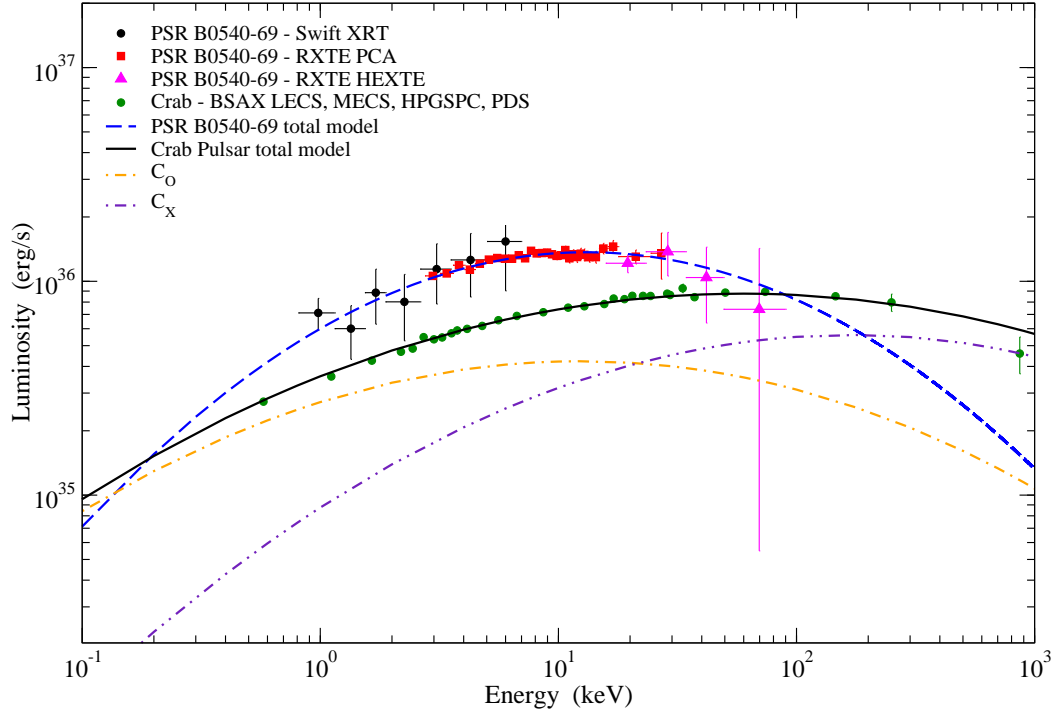} 
\caption{Distance-corrected luminosity for PSR~B0540-69 and Crab. The distance values used are $d=50$ kpc and $d=2$ kpc, respectively. Crab data is from the four NFI onboard BeppoSAX and were taken from Massaro et al. (2006). Also plotted are the log-parabolic model for PSR~B0540-69 (blue dashed line) and the total multicomponent model for Crab (black continuous line, with the $C_{O}$ and $C_{X}$ components as orange dash-dotted line and violet dash-dot-dotted line). See text for discussion.}\label{f:rescaled}
\end{figure*}

Most of the INTEGRAL Galactic PWNs are likely associated with HESS TeV sources (e.g. PSR~J1617-5055 and PSR~J1811-1925).
However, estabilishing the PSR~B0540-69 possible nebular TeV emission is difficult, because of the low declination of this source and, mainly, of the high distance. 
If we assume that the synchrotron nebula powered by PSR~B0540-69 has the same intrinsic TeV brightness of the Crab, the ground flux is reduced to a value of the order of some milliCrab (by a factor $\sim$400). Very long ($>$100 h) dedicated HESS pointings, therefore, are required to detect the possible emission (Hofmann, 2001). 
Another intriguing possibility is that this system lacks an appreciable intrinsic TeV emission. 
It is thought (de Jager \& Djannati-Ata\"i, 2008) that efficient high energy $\gamma$-ray emission is suppressed when the synchrotron losses dominates over the Inverse Compton losses, and this can happen, for example, for high magnetic energy densites or low background radiation fields,
although there is no compelling evidence for both hypotheses. 

Petre et al. (2007), under reasonable assumptions, derive an equipartition magnetic field for the PWN, and found a value of  $B \simeq 8\cdot10^{-4}$ G, higher than that estimated for the Crab by a factor $\sim$2-3 (de Jager \& Harding, 1992), and implying a synchrotron lifetime of electrons of about 3 years.
The efficiency for the very high energy IC emission should therefore be lower with respect to the Crab, making further difficult the detectability of TeV flux.
It will be interesting to see if GLAST will be able to detect nebular emission in the GeV range, and to check if the derived PWN parameters are consistent with the previous estimates.

Note also that an EGRET source, 3EG~J5033-6916, is located about 0.6$^{\circ}$ south-east from the pulsar (thus compatible with the PSF radius), but its association with PSR~B0540-69 is unlikely. This $\gamma$-ray source has been interpreted as due to diffuse emission originating from the interaction between cosmic rays and the LMC interstellar medium (Sreekumar et al., 1992; Lin et al., 1996), and the observed flux is consistent with the theoretical predictions (Fichtel et al., 1991).

\subsection{Pulsed spectrum}

Massaro et al. (2006) developed a phenomenological model of the Crab pulsar X and $\gamma$-ray emission, explained as due to two different emission components $C_{O}$ and $C_{X}$, having a log-parabolic spectrum with the same curvature ($b \simeq 0.16$) but different energy peaks and phase distributions. 
This multicomponent (MC) model also describes the energy variations of the pulse profile.
The $\gamma$-ray emission above 10 MeV is well reproduced, both in the total and phase-resolved spectra, assuming that the  $C_{O}$ and $C_{X}$ have an higher-energy counterpart  $C_{O\gamma}$ and $C_{X\gamma}$, with the same phase distributions and the same log-parabolic spectra at different peak energies.

In Figure \ref{f:rescaled} the pulsed spectra of both PSR~B0540-69 and Crab are plotted, rescaled for the distance values (we assumed $d=50$ kpc and $d=2$ kpc, respectively), together with the $C_{O}$ and $C_{X}$ components of the Crab MC model.
The  emission of PSR~B0540-69 is higher by a factor of about 2 than the Crab, although the true value is strongly dependent on the assumed distance (see Schaefer et al., 2008, for a discussion on the LMC distance value), but also on the difference in the solid angle subtended by the emission beam, here assumed to be isotropic. If we adopt the distance values from HSTKP ($50.1\pm2.4$ kpc, Freedman et al., 2001), the luminosity ratio lies in the range of 1.1--3.7. 

It is interesting to note that, although PSR~B0540-69 spectral curvature is larger ($b \simeq 0.24$ versus $b \simeq 0.16$ for the Crab), the peak energy is sensibly close to the $E_{p} \simeq 12$ keV value of the $C_{O}$ component of the MC model. 
The absence of a variation with energy of the light curve also makes plausible the interpretation that the PSR~B0540-69 emission is only due to its counterpart of the Crab $C_{O}$ component.
The pulse shapes of these pulsars are however very different, and this could be due either to different viewing conditions or to unlike properties of the emission regions.

Takata \& Chang (2007) recently developed an outer-gap model of Crab and PSR~B0540-69 emission, extending the emission region also inside the null-charge surface, along the last closed field line. 
Their results are in broad agreement with the MC phenomenological model, if we identify the $C_{O}$ component as due to the emission beyond the null charge surface to the light cylinder, and the $C_{X}$ component as coming from the region below the null charge surface up to the neutron star surface.
In their model, the X-ray spectrum is due to synchrotron emission from secondary e$^{\pm}$ pairs, and has a curved shape similar to our log-parabolic fit.
Takata \& Chang (2007) explain the different pulse shapes of Crab and PSR~B0540-69 assuming a different inclination angle ($\alpha \simeq 50^{\circ}$ and $\alpha \simeq 30^{\circ}$, respectively) between the spin axis and the magnetic axis, and a different viewing angle $\xi$. Also, the latter pulsar has a thicker emission region of the secondary pairs.

If we assume that also PSR~B0540-69 has a $C_{O\gamma}$ component at the same peak energy (300 MeV) of the Crab equivalent, and with the same relative normalizations of the two log-parabolic spectra, we can obtain a value of the flux at the peak frequency: $\sim 4\cdot 10^{-7}$ MeV~cm$^{-2}$~s$^{-1}$.
This $\gamma$-ray flux prediction is of the same order of magnitude with respect to the one derived from Takata \& Chang (2007) by Inverse Compton scattering.
The predicted pulsed flux is about one order of magnitude lower than the $5 \sigma$ one-year GLAST survey 
sensitivity\footnote{ \url{http://www-glast.slac.stanford.edu/software/IS/glast_lat_performance.htm} }. 
Note that the EGRET upper limit (Thompson et al., 1994) is about $10^{-5}$ MeV~cm$^{-2}$~s$^{-1}$ in the energy range above 100 MeV.


\section{Conclusions}\label{conclusions}
PSR~B0540-69 is an interesting young pulsar, showing many similarities but also some differences with respect to the Crab. 
The study of this pulsar, although made difficult by its high distance and apparent faintness, could help the theoretical understanding of the characteristics of the high-energy emission from the pulsars and their PWNs.

With the analysis of the complete RXTE dataset of observations of PSR~B0540-69, 
exploiting for the spectral analysis the longest gain epoch that spans more than half of these data,
together with that of new SWIFT/XRT and INTEGRAL/IBIS data, we have compiled an accurate and up-to-date picture of the emission from this source in the X and soft $\gamma$-rays, from 0.7 to 200 keV. Thanks to the full HEXTE dataset we have also determined the contribution of pulsed hard X-ray emission to the total one.

Regarding the pulsed emission, we confirm the presence of a curved spectrum, peaking at about 12 keV, well described by a log-parabolic law up to $\ga$100 keV. We have shown also that it presents some similarities with the Crab pulsed emission.

We have studied in detail the total (pulsar+PWN) emission from this source, in the energy range 0.7--100 keV, and we have shown that the major (about 75--80\%) contribution comes from the synchrotron nebula.
The comparison of PSR~B0540-69 with other IBIS pulsar/PWN systems shows that it has the highest conversion efficiency in the hard X/soft $\gamma$-ray band, with similar spectral properties with respect to the other sources. Most of these INTEGRAL PSR/PWN systems have been associated with HESS sources. PSR~B0540-69 looks different, in this respect, not  showing evidence of emission in TeV range.
We have discussed the possible instrumental limitations, due to the low flux and declination, but there is also the possibility that this source genuinely lacks TeV emission.

Future observations, especially in the hard X-ray band, will further refine this picture.
High angular resolution long Chandra X-ray observations will permit an accurate study of the synchrotron nebula to derive its physical parameters.
Additional INTEGRAL pointings will increase the statistics of the total emission, while the advent of hard X-ray focusing telescopes, like Simbol-X (Ferrando et al., 2005), will help to disentangle the pulsed emission from the nebular one also in the HEXTE and IBIS energy range.
Moreover, deep GLAST observations could be useful to estabilish more strict upper limits to the $\gamma$-ray emission from this source.


\section*{Acknowledgments}
We thank Vanessa McBride and Raffaella Landi for the help in data reduction and useful discussions.
We also thank the anonymous referee for interesting comments and useful suggestions.

\label{lastpage}

\end{document}